\documentclass[iicol]{sn-jnl}
\usepackage[T1]{fontenc} 
\usepackage{mathptmx}
\usepackage{graphicx}
\usepackage{dcolumn}
\usepackage{bm}
\usepackage{makecell}
\usepackage{soul}
\usepackage{amsmath}
\usepackage{cite}

\begin{document}

\title[Article Title]{Optical pumping and laser slowing of a heavy molecule}

\author[1]{\fnm{Shuhua} \sur{Deng}}

\author[1]{\fnm{Shoukang} \sur{Yang}}

\author*[1]{\fnm{Zixuan} \sur{Zeng}}\email{zixuanzeng@zju.edu.cn}

\author*[1,2]{\fnm{Bo} \sur{Yan}}\email{yanbohang@zju.edu.cn}

\affil[1]{\orgdiv{Zhejiang Key Laboratory of Micro-nano Quantum Chips and Quantum Control}, \orgdiv{School of Physics and State Key Laboratory for Extreme Photonics and Instrumentation}, \orgname{Zhejiang University}, \orgaddress{\city{Hangzhou}, \postcode{310027}, \country{China}}}

\affil[2]{\orgdiv{College of Optical Science and Engineering}, \orgname{Zhejiang University}, \orgaddress{\city{Hangzhou}, \postcode{310027}, \country{China}}}

\date{\today}

\abstract{Precision measurements of the electron's electric dipole moment (eEDM) are critical for testing fundamental symmetries in particle physics, and heavy polar molecules—such as barium monofluoride (BaF)—have emerged as promising candidates for advancing the sensitivity. However, the achievement of a 3D magneto-optical trap (MOT) required slowing BaF molecules to near-zero velocity by scattering over 10$^4$ photons per molecule, demanding a quasi-cycling transition with minimal leakage. We present a detailed study of the leakage channels, including higher vibrational and rotational states. By combining microwave remixing with optical pumping of rotational and vibrational dark states, we reduced the total leakage fraction to 10$^{-5}$. Using frequency-chirped laser slowing, we slowed a subset of buffer-gas-cooled BaF molecules from approximately 80 $\text{m }\text{s}^{-1}$ to near-zero velocity, which is critical for efficient MOT loading. This work establishes the technical foundation for precision eEDM measurements using laser-cooled heavy molecules.}

\maketitle

\section{Introduction}

Heavy polar molecules, such as ThO, HfF$^+$, YbF, and BaF, exhibit enhanced internal effective electric fields, making them ideal candidates for the electron's electric dipole moment (eEDM) measurements \cite{Carr2009, Safronova2018, Hudson2011,  Aggarwal2018}. Current state-of-the-art eEDM measurements utilize molecular beams \cite{Andreev2018} or trapped molecular ions \cite{Roussy2023}. The molecular beam experiment features an enormous number of molecules and a significant forward velocity, with molecules traversing the interaction region within a few milliseconds. In contrast, the molecular ion experiment can continuously trap molecules in an ion trap, though the Coulomb repulsive effect imposes limitations on the number. However, recent advances in laser cooling suggest that cold molecules could significantly enhance measurement precision by simultaneously having a large particle number and a long interaction time\cite{Tarbutt2013, Aggarwal2018}. Realizing a 3D magneto-optical trapping (MOT) of molecules is the first step towards such a goal, which has been demonstrated with few molecules \cite{Barry2014, Truppe2017_2, Anderegg2017, Collopy2018, Vilas2022}, and recently extended to BaF \cite{Zeng2024}, SrOH \cite{Lasner2024} and AlF\cite{Padilla2025}.

One of the key challenges in achieving a 3D molecular MOT is laser slowing \cite{Barry_2012, Zhelyazkova_2014, Yeo_2015, Hemmerling_2016, Truppe_2017}. Current laser cooling and magneto-optical trapping of molecules typically utilize buffer-gas-cooled beams \cite{Hutzler2012, Truppe2018}. These beams exhibit a forward velocity on the order of 100 $\text{m }\text{s}^{-1}$, while the transverse velocity—largely determined by the geometry of the mechanical beam collimators used in such experiments—yields on the order of 1 $\text{m }\text{s}^{-1}$. While heavy molecules usually possess a small recoil velocity ($\sim$ 3 $\text{mm }\text{s}^{-1}$) due to their longer transition wavelengths and greater mass, transverse cooling remains relatively straightforward, requiring only about a few hundred photons scatter \cite{Shuman2010, Hummon2013, Lim2018, Bu2022, Debayan2020, Augenbraun2020, Baum2020, McNally2020, Carson2022}. In contrast, longitudinal laser slowing is considerably more challenging. Slowing molecules to near-zero velocity necessitates scattering over 10$^4$ photons, demanding cycling transitions with leakage suppression below 10$^{-5}$. This poses a major challenge for eEDM-optimized molecules, where multiple leakage channels often exhibit leakage fractions exceeding this threshold. Therefore, comprehensive identification and repumping of such channels presents a formidable task, especially for newly targeted molecular species \cite{Lasner2022, Frenett2024}.

\begin{figure}[tp]
\centering
\includegraphics[width=0.47\textwidth]{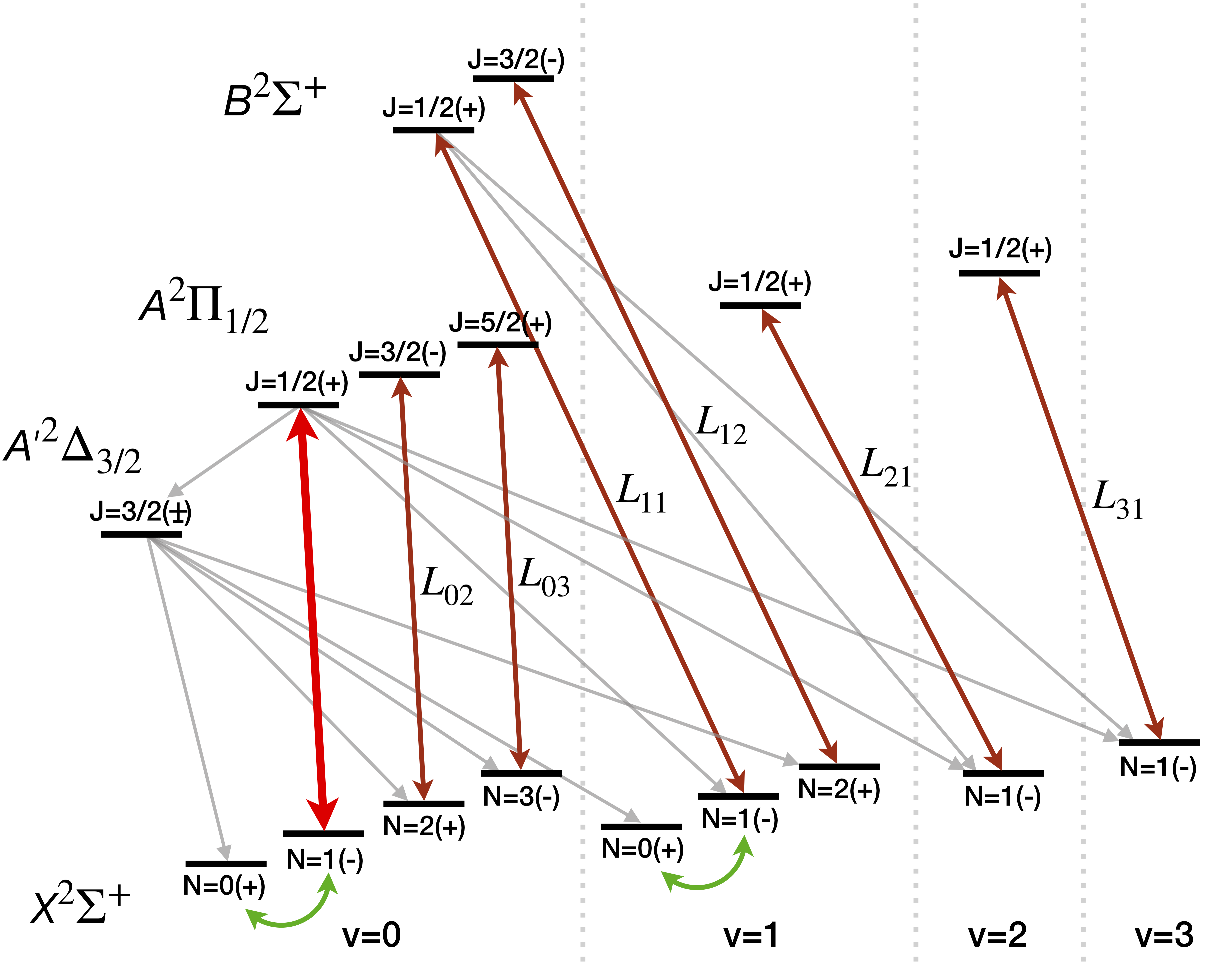}
\caption{
\textbf{The energy levels relevant to the laser slowing of BaF molecules.} The red line denotes the main cooling laser. The brown lines stand for the repump lasers, where the label $L_{vN}$ indicates the vibrational and rotational quantum numbers of the corresponding dark state. 
The green lines denote the microwaves, and the grey lines indicate the principal leakage pathways to higher rotational and vibrational states.}
\label{level}
\end{figure}

\begin{table}[tp]
\renewcommand{\arraystretch}{1.2}
\centering
\begin{tabular}{ccc}
\hline
\hline
~~~~~~~~Label~~~~~~~~ &~~~~~~Dark state~~~~~~~~ & ~~~~~~wavelength (nm)~~~~~~~ \\
\hline
$L_{02}$  &$|v=0,N=2\rangle$  &  $859.8463$  \\
$L_{03}$  &$|v=0,N=3\rangle$  &  $859.8542$  \\
$L_{11}$  &$|v=1,N=1\rangle$  &  $736.7047$  \\
$L_{12}$  &$|v=1,N=2\rangle$  &  $736.7360$  \\
$L_{21}$  &$|v=2,N=1\rangle$  &  $897.9315$  \\
$L_{31}$  &$|v=3,N=1\rangle$  &  $900.1752$  \\
\hline
\hline
\end{tabular}
\caption{\textbf{The primary dark states in the cycling transition of BaF molecules and the corresponding vacuum wavelengths of their repump lasers.} The label $L_{vN}$ indicates the vibrational and rotational quantum numbers of the corresponding dark state. The above wavelengths were measured using a wavemeter (HighFinesse WS7) with an absolute accuracy of 60 MHz.} 
\label{list1}
\end{table}

The BaF molecule is well-suited for laser cooling and trapping. Its electronic transitions within the 700–900 nm spectral range align well with readily available commercial diode laser systems. The energy structure of BaF has been thoroughly characterized by high-resolution spectroscopic studies \cite{Ryzlewicz1982, Barrow1988,Effantin1990, Bernard1992, Guo1995, Steimle2011, Chen_2016, Bu2017, Hao2019, Albrecht2020, Bu2022, Denis2022, Marian2023}. Subsequently,  one-dimensional laser deflection \cite{Chen2017}, laser cooling \cite{Zhang2022, Rockenhaeuser2024} and 3D MOT \cite{Zeng2024} have been successfully demonstrated.  Here, we present a detailed investigation of leakage channels in BaF and corresponding repumping schemes. Using two distinct methods, we identify and precisely measure dark states with leakage fraction exceeding $10^{-5}$. We address these leakages with microwaves and repumping lasers, achieving sufficient photon scattering before decaying to a dark state. Finally, we demonstrate effective deceleration of buffer-gas-cooled BaF molecules using frequency-chirped laser slowing, establishing key techniques for future cold molecule eEDM measurements.

\section{Methods}\label{sec2} 
\subsection{Experimental setup}

Figure \ref{level} shows the energy levels and the relevant transitions in our experiment. The main cooling transition is $|A^2\Pi_{1/2}, v=0, J=1/2,+\rangle\to |X^2\Sigma, v=0, N=1,-\rangle$ with a wavelength of 859.8388 nm. Unlike the dark states emerging from polarization selection rule or coherent population trapping, which could be addressed by magnetic field or polarization modulation, molecules will fall into several stationary dark states through various leakage channels in the process of scattering photons. These leakage channels could be separated into two types. One is to higher vibrational states ($v=1,2,3,\cdots$) characterized by their Franck-Condon factors. The other is to different rotational states due to molecules in the excited $A^2\Pi_{1/2}$ state decaying to the low-lying $A'^2\Delta_{3/2}$ state, which can then decay to different rotational states ($N=0,2,3,\cdots$) \cite{Chen_2016}. All repump lasers are shown in Fig. \ref{level} with brown lines.  We use $L_{vN}$ and $\lambda_{vN}$ to label the repump lasers and their wavelengths associated with the dark state in the ground state $X^2\Sigma$, vibrational state $v$, and rotational state $N$. Their wavelengths are listed in Table \ref{list1}.

\begin{figure}[tp]
\centering
\vspace{2mm}
\includegraphics[width=0.48\textwidth]{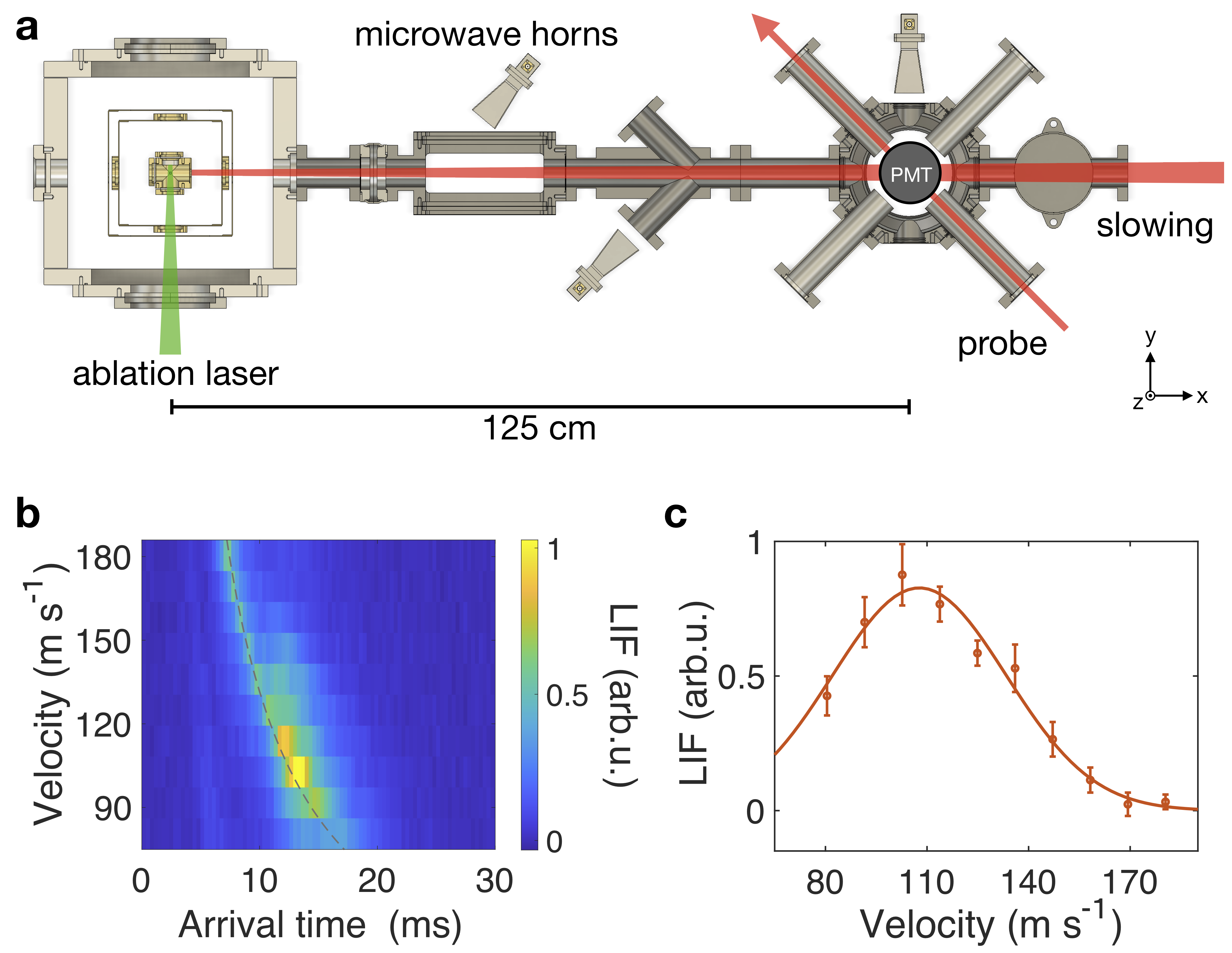}
\caption{\label{setup}
\textbf{Laser slowing experimental setup and molecular velocity distribution.}
(a) The experimental setup consists of a cryogenic buffer gas source and a series of vacuum chambers. Slowing, probe, and ablation lasers are employed in the experiment via different optical accesses. (b) The time-velocity density plot of the light-induced fluorescence (LIF) of BaF molecules at a helium flow rate of 0.5 SCCM. The vertical axis represents molecular velocity, and the horizontal axis represents molecular arrival time at the probe region after laser ablation. The dashed curve is a guideline, depicting arrival times calculated from molecular velocity $v$ and apparatus geometry $t=0.5\text{ ms}+1.25\text{ m}\cdot v^{-1}$ for ease of experimental data comparison and analysis. The orange circles in panel (c) are obtained by integrating data in panel (b) from 5 ms to 20 ms. The orange curve is a Gaussian fitting. The error bars represent the standard error of the mean.
}
\end{figure}

The experimental setup is depicted in Fig. \ref{setup}. A molecular beam is generated within a 4K buffer-gas cell, with approximately $10^6$ molecules in the $|v=0,N=1\rangle$ state reaching the detection region within the MOT chamber\cite{Bu2017}. A probe beam, comprising the main cooling laser with a 4 mm waist diameter ($1/e^2$ intensity diameter) and 1 mW optical power, is oriented at a 45-degree angle across the molecular beam to enable velocity-sensitive detection. The probe beam comprises four frequency components, individually generated by three acousto-optic modulators (AOMs) with separations of $\{0, 17, 124, 152 \}$ MHz, which precisely correspond to four hyperfine transitions of BaF. The natural linewidth of BaF is only $2\pi\times2.8$ MHz, and both the excited states and ground states have large hyperfine splittings ($6-30\Gamma$). This ensures that only molecules with velocities enabling exact four-frequency alignment to the hyperfine structure can undergo cycling transitions, yielding a clean spectrum. While molecules with different velocities may experience crosstalk involving three or two frequencies, such crosstalk cannot sustain cycling transitions. Consequently, the photon count from crosstalk scattering is significantly lower than the target molecular signal we detect. By sequentially scanning its frequency across successive experimental cycles, we obtain time-resolved photomultiplier tube (PMT) signals that capture the entire molecular velocity distribution. This method yields the time-velocity density plot as shown in Fig.~\ref{setup}(b), where the dashed curve serves as a guideline, depicting arrival times calculated from molecular velocities and apparatus geometry for ease of experimental data comparison and analysis. Integrating the signal temporally between 5 and 20 ms yields a velocity distribution that peaks around 110 $\text{m }\text{s}^{-1}$, as shown in Fig. \ref{setup}(c).

\begin{figure}[tp]
\centering
\includegraphics[width=0.45\textwidth]{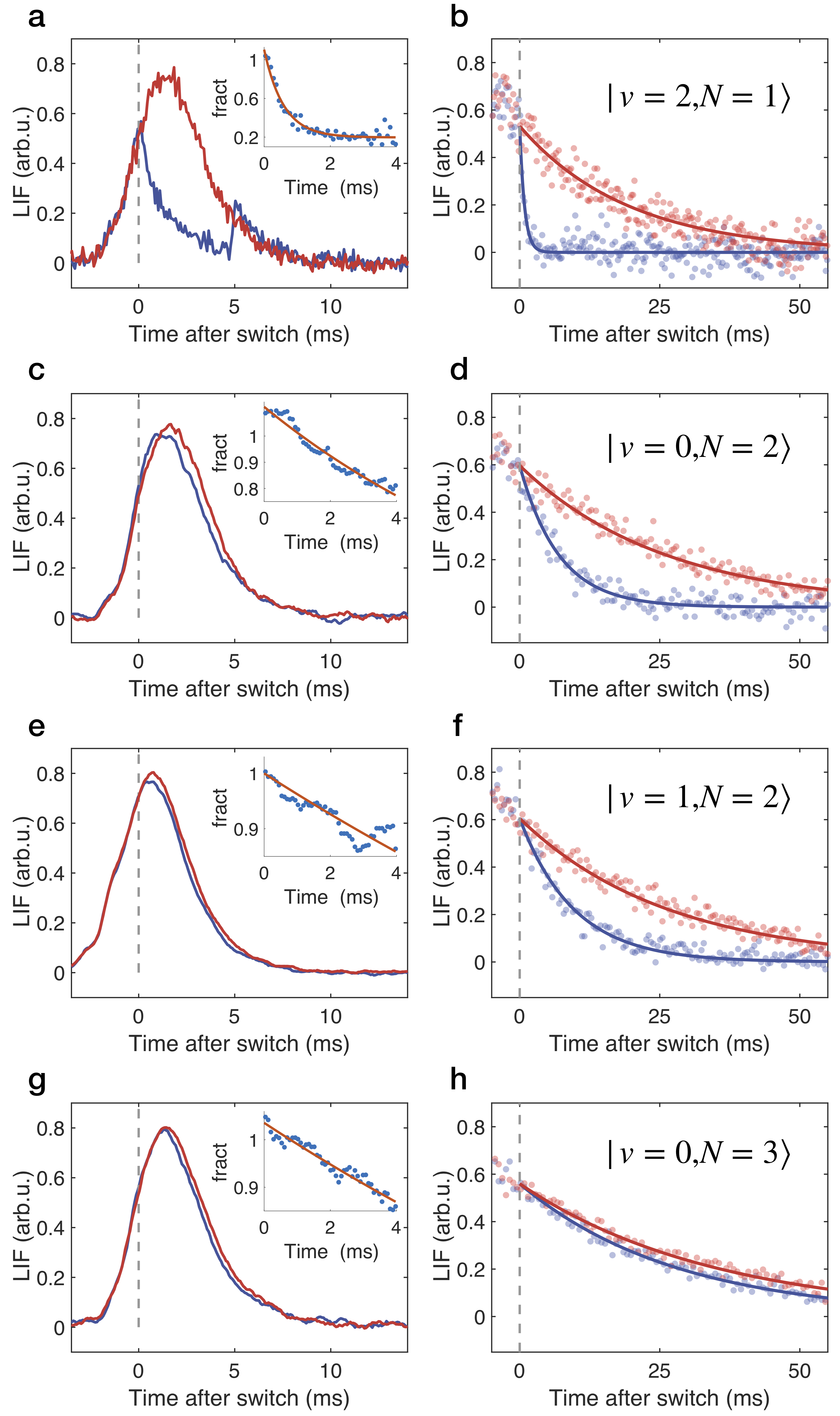}
\caption{\label{leakage}
\textbf{Typical leakage measurement data showing laser-induced fluorescence signals for selected molecular states. The data features laser-induced fluorescence (LIF) signals for the states $|v=2,N=1,-\rangle$(a, b), $|v=0,N=2,+\rangle$(c, d), $|v=1,N=2,+\rangle$(e, f), and $|v=0,N=3,-\rangle$(g, h).} In all the panels, the red curve represents the LIF signal with all repump lasers on, while the blue curve shows the LIF signal with one specific repump laser switched off at $t=0$ ms. The data in panels (a), (c), (e), (g) were obtained via the slowing laser method, with insets displaying the normalized LIF signals fitted with exponential decays. The data in panels (b), (d), (f), (h) were acquired using the magneto-optical trap (MOT) method with exponential decay fits. All data are averages over $\sim200$ experimental cycles. For clarity, panels (c), (e), (g) were smoothed using a moving average with a 0.8 ms window.
}
\end{figure}

A slowing laser, containing the main cooling laser and all the repump lasers, is directed counter-propagating to the molecular beam. The main cooling laser has a power of 200 mW (140 mW) for white light slowing (chirped slowing), $v=1$ repump laser has a power of 130 mW, $v=2$ repump laser has a power of 60 mW, and the power of the remaining repump lasers is on the order of 10 mW.  All lasers have a waist diameter ($1/e^2$ intensity diameter) of 11 mm at the entrance of the viewport and are slightly converged such that their focal point lies approximately 50 cm behind the buffer gas cell. Additional vacuum windows are installed at the rear of both the buffer gas cell and the vacuum chamber, allowing the slowing laser to traverse these components and exit the apparatus. This configuration facilitates beam alignment and minimizes the stray light. In the slowing experiment, Zeeman dark states are remixed by a 5 Gauss magnetic field at a 45-degree angle to the linear polarization of the slowing beam. The fluorescence induced by the slowing laser is collected by a photomultiplier tube (PMT), forming a pulsed slowing laser-induced fluorescence (LIF) signal, as shown by red curves on panels (a), (c), (e), (g) of Fig. \ref{leakage}. An 860 nm bandpass filter is installed to block all fluorescence except 860 nm photons.

\begin{table*}[tp]
\renewcommand{\arraystretch}{1.2}
\centering
\resizebox{\linewidth}{!}{\begin{tabular}{cccc}
\hline
\hline
~~~~Dark state in $X^2\Sigma^+$~~~ & ~~~~~~Main Leakage Pathways~~~~~~ & ~~~~~~~~~~~~~$\eta_\text{S}$~~~~~~~~~~~~~  & ~~~~~~~~~~~~~$\eta_\text{M}$~~~~~~~~~~~~~  \\
\hline
$(v=0\,\&\,1,N=0)$  &  $|A,v=0,J=1/2,+\rangle\to|A',v=0,J=3/2,-\rangle\to|X,v=0,N=0,+\rangle$  & $0.7(4)\times 10^{-4}$  & $0.9(5)\times 10^{-4}$\\
\hline
$(v=0,N=2)$  &  $|A,v=0,J=1/2,+\rangle\to|A',v=0,J=3/2,-\rangle\to|X,v=0,N=2,+\rangle$  & $0.9(4)\times 10^{-4}$  & $1.5(3)\times 10^{-4}$\\
\hline
$(v=0,N=3)$  &  $|A,v=0,J=1/2,+\rangle\to|A',v=0,J=3/2,\pm\rangle\to|X,v=0,N=3,-\rangle$  & $4(2)\times 10^{-5}$  & $1.0(2)\times 10^{-5}$\\
\hline
$(v=1,N=2)$  &  $|A,v=0,J=1/2,+\rangle\to|A',v=0,J=3/2,-\rangle\to|X,v=1,N=2,+\rangle$  & $4(1)\times 10^{-5}$  & $9(2)\times 10^{-5}$\\
\hline
$(v=3,N=1)$  &   \makecell{$|A,v=0,J=1/2,+\rangle\to|X,v=3,N=1,-\rangle$ \\ $|B,v=0,J=1/2,+\rangle\to|X,v=3,N=1,-\rangle$  } & $2(1)\times 10^{-5}$  & $5(1)\times 10^{-5}$\\
\hline
\hline
\end{tabular}}
\caption{
\textbf{Leakage fractions of dark states measured via slowing laser or magneto-optical trap method.} The primary leakage pathways are listed. $\eta_\text{S}$ and $\eta_\text{M}$ represent the leakage fractions measured by the slowing laser method and the magneto-optical trap (MOT) method, respectively. All the leakage fractions, except $\eta_\text{S}$ for the $|v=3, N=1\rangle$ state, are derived from Eq. (\ref{Eq1}). The $\eta_\text{S}$ for the $|v=3, N=1\rangle$ state is estimated by measuring the revival rate of the laser-induced fluorescence(LIF) signal when the repump laser $L_{31}$ is switched off.
}
\label{list2}
\end{table*}

\subsection{Leakage measurement}

The leakage fraction of a dark state can be determined by measuring the LIF signal upon switching off the repump laser. Typically, the LIF signal with all repump lasers on also has an exponential decay behavior with a $1/e$ lifetime $\tau_\text{repump on}$. By contrast, switching off one repump laser causes the LIF signal to exhibit a much faster exponential decay with a $1/e$ lifetime $\tau_\text{repump off}$ due to the leakage into a dark state. If molecules scatter photons at a rate of $R$, the leakage fraction $\eta$ can be calculated as
\begin{equation}\label{Eq1}
\eta=(\frac{1}{\tau_\text{repump off}}-\frac{1}{\tau_\text{repump on}})\frac{1}{R}.
\end {equation}

We first derived the scattering rate $R$ from Eq. (\ref{Eq1}) using a known theoretical leakage fraction with high calculation accuracy \cite{Barry2014, Truppe2017_2, Williams_2017}, then determined all the other unknown leakage fractions from Eq. (\ref{Eq1}) via this scattering rate.

Two methods are employed to measure the LIF decay: one using the slowing laser and the other using a MOT. We label the corresponding scattering rates as $R_\text{S}$ and $R_\text{M}$, and the resulting leakage fractions as $\eta_\text{S}$ and $\eta_\text{M}$, respectively.

In the slowing laser method, the slowing LIF signal $f_\text{repump off}(t)$ is recorded with one repump laser turned off at t = 0 ms, as shown in Fig. \ref{leakage}. To account for the temporal variation of the LIF signal due to all factors except dark state leakage, we normalize the signal to a reference signal $f_\text{repump on}(t)$ recorded without switching off the repump laser. Because molecules are flying, the LIF signals exhibit complex features and cannot be fitted well with single exponential decay curves. Instead of using Eq. (\ref{Eq1}), we use the normalized LIF signals $f_\text{normalized}(t)=f_\text{repump off}(t)/f_\text{repump on}(t)$, as shown in the insets, and fit them with an exponential decay function to extract the decay time $\tau_\text{S}$, and then determine the leakage fraction $\eta_\text{S}=1/\tau_\text{S} R_\text{S}$ accordingly \cite{Yeo_2015}.

In the MOT detection method, we measure the MOT LIF signal with one repump laser switched off after MOT loading, as shown on panels (b), (d), (f), (h) of Fig. \ref{leakage}. Notably, the MOT LIF signal decays even with all repump lasers on, limited by factors such as background vacuum pressure, residual leakages, and potential technical issues. The lifetimes of these two LIF signals are respectively obtained via exponential fitting. The leakage fraction $\eta_\text{M}$ is then calculated according to Eq. (\ref{Eq1}). Because molecules are trapped, this method offers a more accurate measurement. However, this method is only applicable after achieving a molecular MOT, which is not always feasible. 
 
Typical data from these two types of measurement are shown in Fig. \ref{leakage}. The summary of all measurements is listed in Table \ref{list2} and will be discussed in detail below.

\section{Results}
\subsection{Vibrational leakages}
The dominant vibrational leakage channel is to $v=1$ via the $|A^2\Pi_{1/2},v=0\rangle \rightarrow |X^2\Sigma,v=1\rangle$ transition, exhibiting a vibrational branching fraction of approximately $4\%$. To mitigate this, we implement optical repumping through the $|B^2\Pi_{1/2}, v=0\rangle\leftarrow|X^2\Sigma, v=1\rangle$ transition at $\lambda_{11} = 737\,\text{nm}$. This transition is strategically chosen to avoid sharing excited states with the main cooling transition, thereby enhancing the net cooling force \cite{Anderegg2017}. This approach has been previously validated in our Doppler cooling experiments \cite{Zhang2022}.

Secondary vibrational leakage to $v=2$ originates from both $|A^2\Pi_{1/2},v=0\rangle{\rightarrow}|X^2\Sigma^+,v=2\rangle$ and $|B^2\Sigma,v=0\rangle{\rightarrow}|X^2\Sigma^+,v=2\rangle$ transitions, with a theoretical leakage fraction of $1.6{\times}10^{-3}$ (See Supplementary Note 1). When the scattering rate $R$ is known, the leakage fraction can be determined by Eq. (\ref{Eq1}). However, independent measurement of $R$ brings a relatively high degree of uncertainty. Considering both the measurement accuracy of the LIF signal and the calculation accuracy of the Franck-Condon factor, we use this theoretical leakage fraction to $v=2$ as a reference for determining the scattering rate of the cycling transition. In our experiment, we obtain $\tau_\text{S} = 0.59(7)\ \text{ms}$  and $\tau_\text{M} = 0.9(1)\ \text{ms}$, yielding $R_\text{S}=1.1(2)\times10^6$ s$^{-1}$ and $R_\text{M}=0.7(1)\times10^6$ s$^{-1}$. This scattering rate corresponds to a deceleration rate of around 3300 $\text{m }\text{s}^{-2}$ during the white light slowing.

\subsection{Rotational leakages}
The metastable $A'^2\Delta$ state of BaF lies 27~THz below the $A^2\Pi$ excited state, with a leakage fraction of around $10^{-4}$. Thus, excited molecules undergo $A{\rightarrow}A'{\rightarrow}X$ cascade decays, which invert the initial parity. This leads to population in $N{=}0$ and $N{=}2$ rotational states (parity $+$), rather than returning to the initial $N{=}1$ state (parity $-$). Additional rotational leakage could be induced by a magnetic dipole transition which connects two states with the same parity. Some strong magnetic dipole transitions have been observed experimentally\cite{Yang1999, Moritz2012}. Because parity selection rules prevent direct optical pumping from $N=0$ or $N=2$ to $N=1$, we employ a hybrid microwave-optical approach\cite{Yeo_2015, Norrgard2016, Collopy2018}. As depicted in Fig.~\ref{level}, Microwave fields remix hyperfine states within the $N{=}0$ and $N{=}1$ manifolds \cite{Zeng2024}. The microwave frequencies are generated by independent microwave sources, each amplified by a 26 dBm power amplifier, combined together by a power combiner and then divided into four channels: three channels deliver approximately 10 dBm of power (per frequency) to each of the three microwave horns shown in Fig. \ref{setup}, while the fourth channel is used for monitoring. Because this scheme increases the number of ground states involved in the cycling transition, it results in a weaker slowing force compared to the all-optical pumping scheme.

Rotational state repumping for $X^2\Sigma(v=0,N=2)$ is achieved via the $|A^2\Pi,v=0,J=3/2,-\rangle\leftarrow |X^2\Sigma,v=0,N=2,+\rangle$ transition at $\lambda_{02} = 859.8463$ nm (Table~\ref{list1}). Figures~\ref{leakage}(a) and \ref{leakage}(b) demonstrate the LIF decay curves to $X^2\Sigma(v=0,N=2)$ with two different methods. They yield $\eta_\text{S}=0.9(4){\times}10^{-4}$ and $\eta_\text{M}=1.5(3){\times}10^{-4}$. Similar experiments are done by switching off the 13 GHz microwave while keeping the $\lambda_{02}$ laser continuously on. Because the molecules falling into the $X^2\Sigma(v=0, N=2)$ dark state need to be pumped to the $X^2\Sigma(v=0, N=0)$ state first and then rejoin the cycling transition by microwave remixing, the leakage fraction measured by switching off the 13 GHz microwave represents the combined leakage of both $X^2\Sigma(v=0, N=0)$ and $X^2\Sigma(v=0, N=2)$ dark states: $\eta_\text{S}=1.6(4)\times10^{-4}$ and $\eta_\text{M}=2.4(5)\times10^{-4}$.

\subsection{Higher-order leakages}
While the aforementioned repumping schemes achieve leakage suppression at the $10^{-4}$ level, reaching the $10^{-5}$ threshold requires addressing higher-order leakage channels. 
First, there are weak decays from $|A^2\Pi_{1/2},v{=}0,J{=}1/2,+\rangle$ and $|B^2\Sigma,v{=}0,J{=}1/2,+\rangle$ to $|X^2\Sigma,v{=}3,N{=}1,-\rangle$, we repump it with a  $\lambda_{31}=900.1752$ nm laser. Second, kHz-level splitting between $\Delta$-state parity manifolds enables residual electric field-induced mixing, populating $|X^2\Sigma,v{=}0,N{=}3,-\rangle$ through $A'{\rightarrow}X$ transitions. We address it with a $\lambda_{03} = 859.8542$ nm repump laser. Third, non-diagonal Franck-Condon factors ($\sim10\%$ $v'{=}0{\rightarrow}v{=}1$ branching fraction) in $\Delta{\rightarrow}\Sigma$ transitions create $|X^2\Sigma,v{=}1,N{=}0,2,+\rangle$ populations. Additionally, the repump laser $L_{02}$ introduces further leakage. This is because the repumping efficiency for rotational states is lower compared to that for vibrational states. The $A^2\Pi_{1/2}(v=0, J=3/2,-)\leftarrow X^2\Sigma(v=0, N=2,+)$ transition is a semi-cycling transition. According to the rotational branching ratio, only 17\% molecules are pumped to the $X^2\Sigma(v=0, N=0, +)$ state after an excitation, while 83\% molecules fall back to the $X^2\Sigma(v=0, N=2, +)$ state. When using this transition to repump the rotational state, multiple photon scatterings are required to drive the molecules back to the $X^2\Sigma(v=0, N=0, +)$ state. As a result, this process increases the leakage fraction of the $X^2\Sigma(v=1, N=2, +)$ state via a new leakage pathway, namely $A^2\Pi_{1/2}(v=0, J=3/2, -)\to X^2\Sigma(v=1, N=2, +)$. We mitigate this using a combination of microwave remixing of $\{v{=}1; N{=}0\leftrightarrow N=1\}$ and a $\lambda_{12}=736.7360$ nm repump laser. These repump lasers are illustrated in Fig. \ref{level}, with wavelengths listed in Table \ref{list1}. Representative measurement data are presented in Fig. \ref{leakage} and the measured leakage fractions are listed in Table \ref{list2}. 

As can be seen from Table \ref{list2}, the leakage fractions of these dark states measured by the slowing laser and MOT methods are generally in agreement with each other. The uncertainty of the measurements is mainly derived from the uncertainty of the scattering rate $R$. A rather obvious difference lies in the leakage fractions of the $|X^2\Sigma,v=0, N = 3\rangle$ state in the laser slowing process and in the MOT. Since the leakage of the $|X^2\Sigma,v=0, N = 3\rangle$ state is caused by the parity mixing effect, residual electric fields of different magnitudes will result in different leakage fractions. Therefore, the difference between the two sets of experimental results may originate from the difference in the residual electric fields in the deceleration path and the MOT chamber.
\begin{figure}[tp]
\centering
\includegraphics[width=0.4\textwidth]{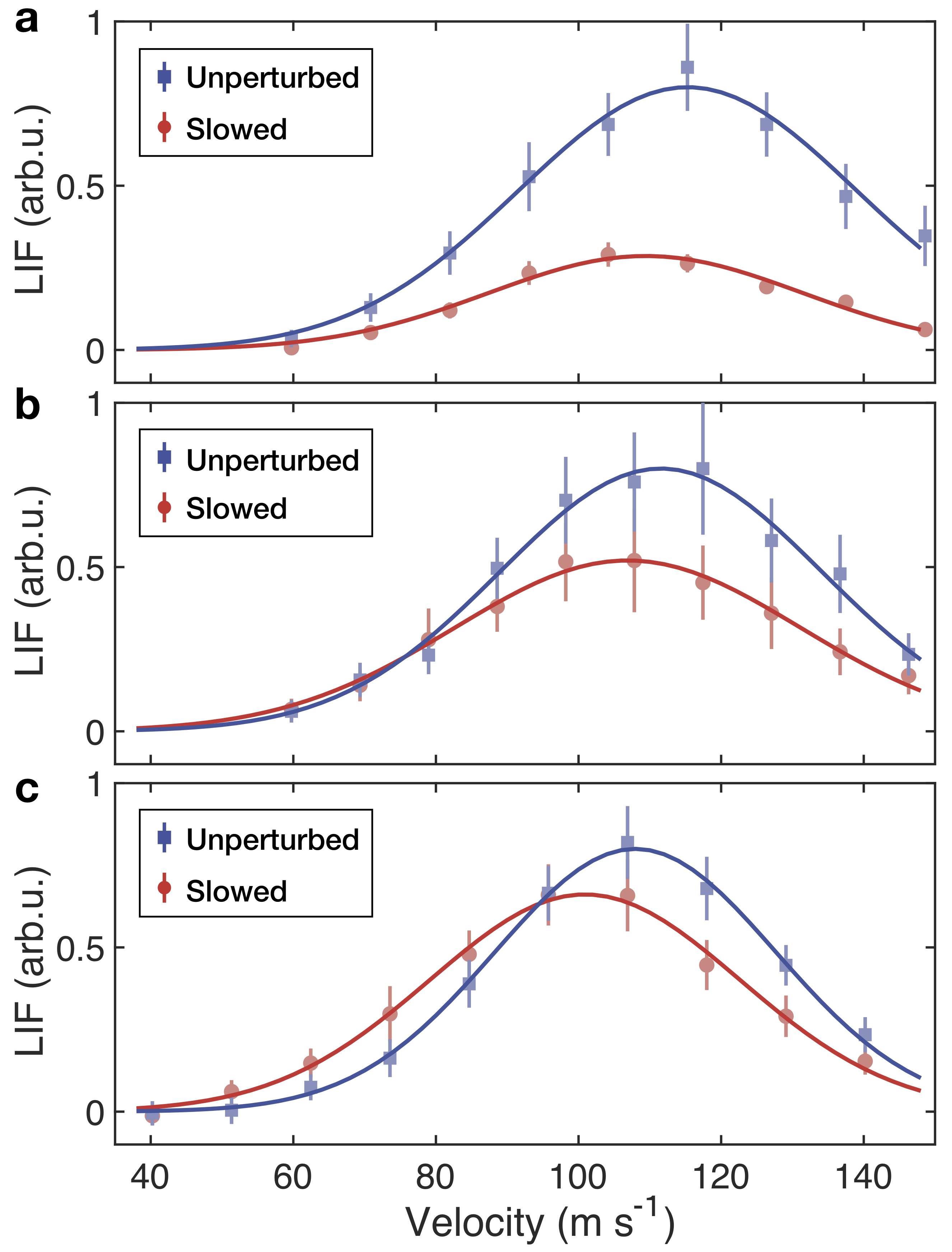}
\caption{
\textbf{Experimental results of laser slowing.} The red circles (blue squares) represent the intensity of light-induced fluorescence (LIF) of slowed (unperturbed) molecular beams at different velocities. The solid curves are Gaussian fittings. (a) White light slowing starts at 0 ms and ends at 6.5 ms in the absence of both microwave remixing and higher-order optical repumping. (b) White light slowing starts at 0 ms and ends at 6.5 ms with microwave remixing and higher-order optical repumping. (c) Chirped slowing from 110 $\text{m }\text{s}^{-1}$ (-128 MHz) to 65 $\text{m }\text{s}^{-1}$ (-76 MHz) in 11 ms with microwave remixing and higher-order optical repumping. The error bars represent the standard error of the mean.}

\label{Slowing1}
\end{figure}

\subsection{Laser slowing of BaF molecule beam}

\begin{figure}[tp]
\centering
\includegraphics[width=0.4\textwidth]{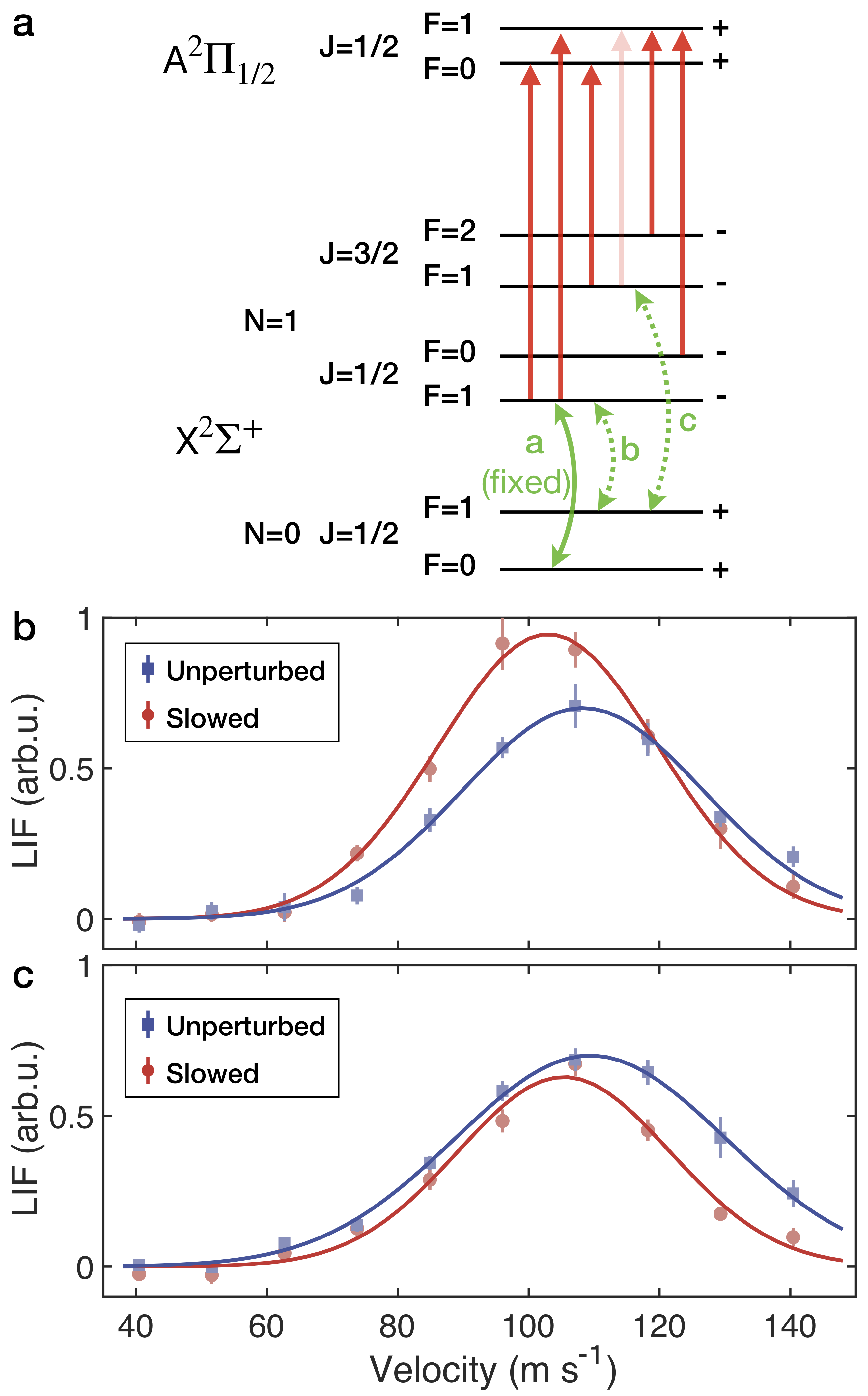}
\caption{
\textbf{Microwave remixing schemes and laser slowing data.} (a) shows the hyperfine structure of the BaF molecule. Red arrows represent possible transitions between $X(N=1,-)$ and $A(J=1/2,+)$ states. Green arrows represent microwave transitions between $X(N=0,+)$ and $X(N=1,-)$ states. Here, the arrow for $A(J=1/2,F=1)\leftarrow X(N=1,J=3/2,F=1)$ is plotted in light red because this transition is nearly forbidden. Arrow a is labelled as ``fixed'' because the microwave frequency it corresponds to remains activated throughout all experimental runs. Two distinct microwave remixing schemes are used in the experiments: Slowing results with frequency components a and b activated are shown in panel (b); those with frequency components a and c activated are shown in panel (c). The frequencies of microwaves a, b, and c are 12.906 GHz, 12.838 GHz, and 12.953 GHz, respectively. The red circles (blue squares) represent the intensity of light-induced fluorescence (LIF) of slowed (unperturbed) molecular beams at different velocities. The solid curves are Gaussian fittings. The frequency of the slowing laser is chirped from a velocity corresponding to 140 $\text{m }\text{s}^{-1}$ (-163 MHz) to 90 $\text{m }\text{s}^{-1}$ (-105 MHz) in 10 ms. The error bars represent the standard error of the mean.
}
\label{Slowing2}
\end{figure}

Two prevalent laser slowing techniques are white light slowing and chirped slowing \cite{Letokhov1976, Prodan1984, Barry_2012, Yeo_2015, Hemmerling_2016, Truppe_2017}. Our first endeavor to decelerate BaF molecules employed the white light slowing method. To achieve white light broadening, we modulate the slowing laser with a 38 MHz electro-optic modulator (EOM) to cover the hyperfine structure with a modulation depth of $M_\text{mod}=2.6$ and a 5.4 MHz EOM for further broadening with $M_\text{mod}=3.2$. Consequently, this led to a frequency width of roughly 200 MHz.

The initial experimental results are presented in Fig. \ref{Slowing1}(a) and (b). In Fig. \ref{Slowing1}(a), since the rotational dark states are not addressed, the total leakage is on the order of $10^{-4}$. {The optimized 6.5 ms} laser slowing process decreases the center velocity by approximately 6 $\text{m }\text{s}^{-1}$; however, it leads to a loss of over 50\% of the molecules. Addressing the higher order leakages (reducing the residual leakage to the order of $10^{-5}$) significantly increases the number of surviving molecules by a factor of $\sim2$, as depicted in Fig. \ref{Slowing1}(b). To preclude overestimating the slowing effect, considering that the rotational repump lasers can augment the number of molecules in the cycling transition, we applied the repump lasers under the same conditions to both the slowed and unperturbed molecular beams. This experimental result demonstrates the importance of minimizing leakages. Nevertheless, the number of slowed molecules is still less than that of unperturbed molecules, thus failing to achieve the primary goal of laser slowing. 

To enhance the efficiency of laser slowing, we make a transition from the white light slowing method to a chirped slowing scheme. Chirped slowing protects low-velocity molecules from the slowing laser and allows them to accumulate in a particular velocity range. We utilize a two-frequency slowing laser with a frequency separation of 134 MHz to address the hyperfine structure of BaF, which gives a narrower deceleration force profile compared to that achieved by using a 38 MHz EOM with $M_\text{mod}=1.6$. By chirping the detuning of the slowing laser from 110 $\text{m }\text{s}^{-1}$ to 65 $\text{m }\text{s}^{-1}$ over 11 ms, we observed an increase in the number of molecules within the 40 $\text{m }\text{s}^{-1}$ to 90 $\text{m }\text{s}^{-1}$ range, as shown in Fig. \ref{Slowing1}(c).

\begin{figure}[tp]
\centering
\includegraphics[width=0.4\textwidth]{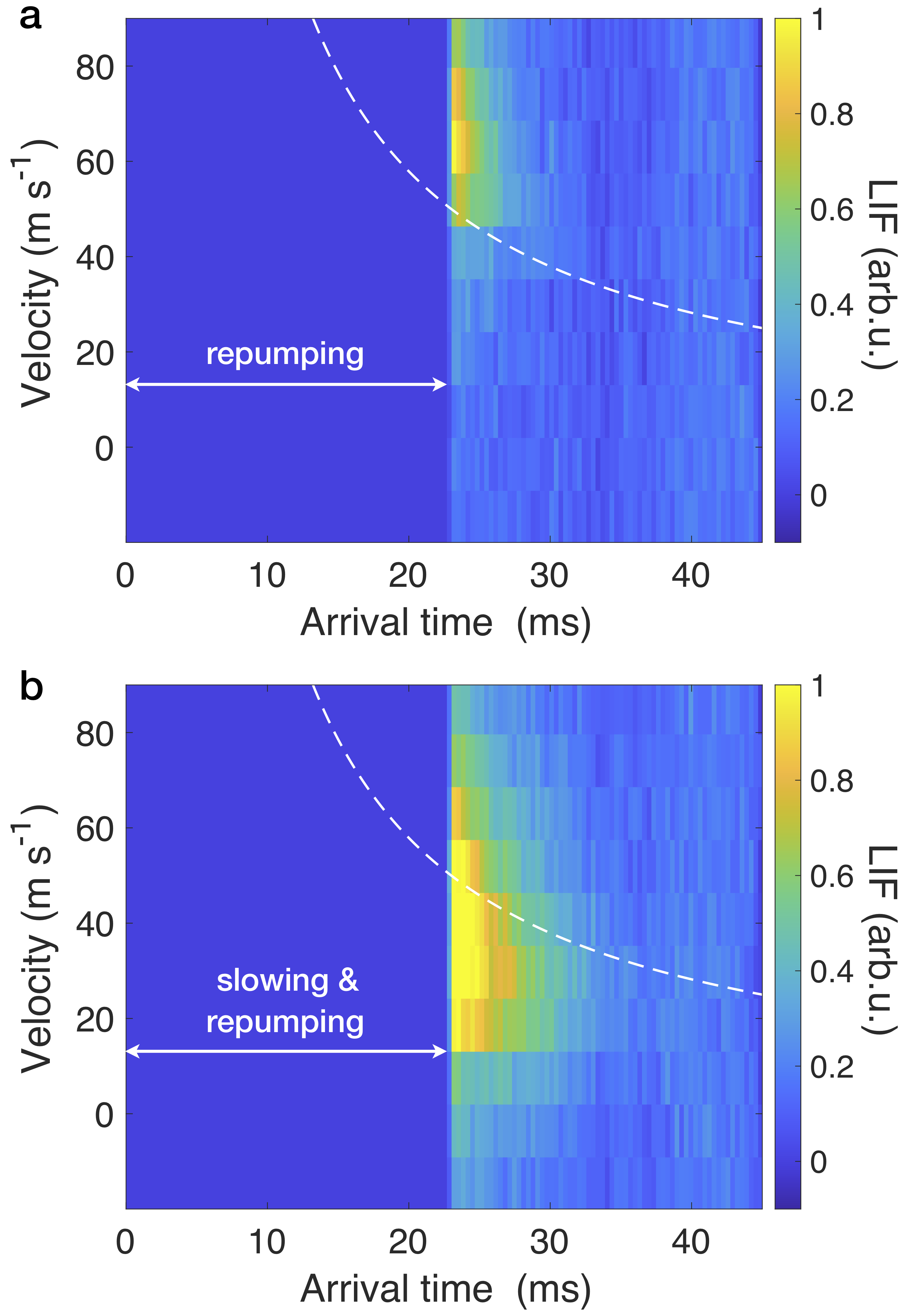}
\caption{
\textbf{Experimental results for the optimal laser slowing.} The figure presents time-velocity distributions of light-induced fluorescence (LIF) for an unperturbed molecular beam (a) and a slowed molecular beam (b). The vertical axis represents molecular velocity, and the horizontal axis represents molecular arrival time at the probe region after laser ablation. The dashed curve is a guideline, depicting arrival times calculated from molecular velocity $v$ and apparatus geometry $t=1\text{ ms}+1.1\text{ m}\cdot v^{-1}$ for ease of experimental data comparison and analysis. The frequency of the slowing laser was linearly chirped from 80 $\text{m }\text{s}^{-1}$ to 10 $\text{m }\text{s}^{-1}$ throughout 21 ms. Since the fluorescence induced by the slowing laser is much stronger than the fluorescence excited by the probe beam, to eliminate its influence on the velocity detection, we started the velocity detection only after turning off the slowing laser at 22 ms.
}
\label{Slowing3}
\end{figure}

Additional experiments have revealed that the choice of microwave remixing scheme plays a crucial role in achieving efficient laser slowing. As illustrated in Fig.~\ref{Slowing2}(a), the $N=0$ manifold comprises two hyperfine states ($F=0$ and $F=1$), and $N=1$ manifold contains four. Consequently, a variety of remixing schemes are feasible. In our experimental investigations, we tested different microwave remixing schemes and observed a substantial reduction in the slowing efficiency when the state $|N=1, J=3/2, F=1\rangle$ was involved in the remixing process. Fig. \ref{Slowing2} (b) presents slowing data obtained using the remixing schemes \{$|N=0, F=0\rangle\leftrightarrow |N=1, J=1/2, F=1\rangle$\} and \{$|N=0, F=1\rangle\leftrightarrow |N=1, J=1/2, F=1\rangle$\}. In contrast, Fig. \ref{Slowing2}(c) demonstrates the slowing performance achieved with the remixing schemes \{$|N=0, F=0\rangle\leftrightarrow |N=1, J=1/2, F=1\rangle$\} and \{$|N=0, F=1\rangle\leftrightarrow |N=1,J=3/2, F=1\rangle$\}. The latter combination led to notably worse laser slowing results.

We attribute this observation to the difference in transition strengths among hyperfine states. The $J$-mixing effect weakens the transition between $X^2\Sigma(N=1,J=3/2,F=1)$ and $A^2\Pi_{1/2}(J=1/2,F=1)$, rendering it nearly forbidden. As a result, molecules in $X^2\Sigma(N=1,J=3/2,F=1)$ are preferentially excited to $A^2\Pi_{1/2}(J=1/2,F=0)$, which has only one Zeeman sublevel. This limitation on the number of accessible excited states in the cycling transition ultimately diminishes the deceleration force, thereby causing the degraded slowing performance as evidenced in Fig. \ref{Slowing2}(c).

After completing all the optimizations, we adjusted the detuning to minimize the final velocity during frequency chirping to obtain BaF molecules with near-zero velocity. The slowing laser is gently broadened by a 12.4 MHz EOM with a modulation depth of $M_\text{mod}=1.7$. The length of the slowing distance has been shortened from 1.25 m to 1.1 m. Moreover, we decreased the helium gas flow rate from 0.5 SCCM to roughly 0.2 SCCM, intending to generate a molecular beam with a lower initial velocity. The ultimate experimental results are presented in Fig. \ref{Slowing3}. The frequency chirping of the slowing light starts at 80 $\text{m }\text{s}^{-1}$ and ends at 10 $\text{m }\text{s}^{-1}$ for 21 ms.  

As depicted in Fig. \ref{Slowing3}(a), the unperturbed BaF molecules are mainly distributed above the dashed line. Most of these molecules had already traversed the detection region before the initiation of velocity detection. In contrast, Fig. \ref{Slowing3}(b) illustrates the slowing result. The majority of the slowed BaF molecules were distributed below the dashed line and displayed a later arrival time when compared to the unperturbed molecules. This result serves as a good starting point for the realization of a molecular MOT.

\section{Conclusions}\label{sec5}

In conclusion, we have identified and measured all the vibrational and rotational leakages in BaF molecules, where the leakages exceed $10^{-5}$. The microwave-optical hybrid pumping scheme, which takes advantage of the large dipole moment characteristic of polar molecules,  effectively mitigates the rotational leakage issue. In detail, microwaves are employed to mix the $N=0$ and $N=1$ states, while optical pumping returns the $N=2$ state to $N=0$. This process minimizes the number of ground states participating in the cycling transition, consequently yielding a saturated scattering rate of $\Gamma/5$ rather than $\Gamma/10$. Due to the susceptibility of the $\Delta$ state to mixing with minute electric fields (on the order of 0.1 V/m), an alternative rotational repumping scheme is proposed: leveraging a 933 nm $A^\prime\Delta(J=3/2,\pm)\rightarrow X^2\Sigma(N=0)$ transition to optically pump molecules in the $X^2\Sigma(N=0)$ state back to the cycling transition. This method further isolates the $X^2\Sigma(N=0)$ state from the $X^2\Sigma(N=1)$ state, thereby enhancing the total scattering rate by $\sim25\%$. Moreover, the 13 GHz microwave wavelength (2.3 cm) is comparable to the dimensions of the vacuum chamber and optical windows, leading to complex diffraction patterns that produce non-uniform intensity distributions or even dark zones within the chamber. Replacing microwave addressing with optical addressing mitigates these issues, ensures homogeneous excitation across the molecular beam, and is worthy of exploration. Using the microwave-optical hybrid pumping scheme, we have successfully achieved laser slowing of a BaF molecular beam, decelerating a subset of molecules from 80 $\text{m }\text{s}^{-1}$ to nearly zero velocity, which paves the way for its capture within a molecular MOT.

\bibliographystyle{sn-nature}
\bibliography{slowing}

\section*{Data availability}
The data that support the findings of this study are available at Science Data Bank\cite{Dataset}.




\end{document}